# Experimental and simulated Performance of Hot Mirror Coatings in a Parabolic Trough Receiver

V.S Kaluba[a,c], Khaled Mohamad[a], P. Ferrer[a,b]

[a]*School of Physics, Materials for Energy Research Group, University of the Witwatersrand, Johannesburg, 2001, South Africa.*

[b]*School of Physics, Mandelstam institute for Theoretical Physics, University of the Witwatersrand, Johannesburg, 2001, South Africa.*

[c]*School of Engineering, University of Zambia, Lusaka, 10101, Zambia*

Thermal radiation is the dominant heat loss mechanism for receiver units on a parabolic solar collector plant at high temperatures. Reduction of these losses is traditionally achieved through the use of an optically selective coating on the absorber pipe, which absorbs visible light well but emits poorly in the IR region. Another possibility is the use of a hot mirror coating on the glass cover of the receiver, which reflects thermal radiation back onto the absorber pipe for reabsorption. In this paper, novel experimental results of a receiver unit operating with a hot mirror coating is presented, and the results between a developed model and a simulation are compared. It is seen that the correspondence is encouragingly close (Chi squared test p-values between 0.995 and 0.80), where the simulation underestimates the experimental performance. Further, simulations to investigate the performance of various candidates for hot mirror coating (ITO, Gold and Silver) in a solar through receiver are presented, where it is seen that the hot mirror coating has access to higher temperature regions (above 700K). Lastly, simulation of effects of variation of some optical parameters on overall plant efficiency, and comparison to existing selective coatings, is presented.

## 1. Introduction

Parabolic trough collectors (PTC) are amongst the most mature commercially developed solar power technologies [1]. They operate by focusing solar radiation along a line onto a receiver unit which exchanges heat with a circulating heat transfer fluid (HTF). The heat gained by the HTF can be used either directly or for electricity production [2]. The receiver traditionally consists of an absorber pipe coated with a selective coating and covered by a glass envelope. The space between the absorber pipe and the glass envelope is evacuated to reduce convective and conductive heat transfer, making radiation the dominant heat loss mechanism.

Numerous works have investigated the thermal properties and heat retention of the receiver unit from various perspectives. Some concentrate on improving the uniformity of thermal distribution of the metal absorber pipe to reduce thermal stress and deformation [3]. Bending of the absorber pipe and breaking of the glass cover have been attributed to the circumferential temperature gradients in the absorber tube. As such, means to reduce these temperature gradients and temperature peaks have been sought to help increase the life span of the receiver. The maximum temperature gradient for safe operation of receiver tubes is about 50 K [4]. Most of these studies either sacrifice the pressure drop of the receiver unit or increase the quality of the absorber pipe and/or other components [3]. Other studies suggest applying inserts into the absorber pipe such as: metal foam, porous disc, perforated plate, or coiled wire turbulators inserts. A metal foam insert reduces the thermal stress, it decreases the temperature difference on the outer surface of the absorber pipe by about 45%. On the other hand, the flow resistance is increased [4]. Experimental [5] and theoretical work [6] was conducted for the porous disc insert application. It showed a significant improvement of the performance of the parabolic trough collector. The perforated plate insert increases the thermal efficiency between 1.2% and 8% according to the numerical study [7]. The coiled wire turbulators insert application has been examined experimentally and numerically [8]. At the pitch distance 30 mm of coiled wire turbulator, the heat transfer enhancement is approximately twice that of the smooth tube [8].

Some studies focus on geometrical structure improvement for the absorber pipe of the receiver unit such as a dimpled tube, unilateral mit-longitudinal vortex-enhanced tube, and symmetric and asymmetric outward convex corrugated tubes. In a numerical study, the overall performance of a dimpled absorber pipe under non uniform heat flux is better than that under uniform heat flux [9]. In the case of unilateral mit-longitudinal vortex-enhanced tube, the overall heat transfer performance is better than the smooth pipe under a wide range of influence parameters of the working conditions and geometric factors [10]. The introduction of symmetric [11] and asymmetric [12] outward convex corrugated tubes can effectively decrease the thermal strain and enhance the heat transfer performance. Researchers have investigated using Nano fluids to enhance the heat transfer, with suspended nanoparticles. The most used nanofluids contain the following nano particles Al, $Fe_2O_3$, $Al_2O_3$, Cu, $TiO_2$, and $SiO_2$ [13][14]. The main use of the nanofluids is to increase the thermal conductivity of the heat transfer fluids. The thermal conductivity increases by decreasing the particle size, and increasing the volume fraction and temperature [14]. E. Bellos found that the use of the nanofluids increases the efficiency of the collector by 4.25% [13].

In PTCs, the dominant heat losses at high temperatures are due to the thermal emission (IR) from the receiver pipe. Conventionally, IR losses are minimized by painting the receiver pipe with a spectrally selective coating, a dielectric film that absorbs well in the visible region of the solar spectrum and emits poorly in the IR region. Much work has been published on selective coating and their properties [15]. This coating typically is not stable chemically and thermally above 500 °C [16] [17], reducing the temperature ceiling, and hence thermal efficiency. The world's most advanced solar receiver tube (Archimede Solar Energy (ASE) [18]) with selective coating operates at temperature up to 580°C [18] with molten salts as Heat Transfer Fluid.

An alternative to the selective coating is coating the inside glass cover around the absorber with a dielectric material that is transparent to the visible region of the solar spectrum and reflects well in the IR region. A coating of this type is referred to as a "hot mirror", and was first implemented for energy efficient windows in automobiles and buildings [16] and for applications related to concentrating photovoltaics and thermophotovoltaics [19][20]. For PTCs, effects of the hot mirror film have been modeled and studied previously. Grena [21] simulated the system including heat reflection using hot mirror films with simplifying assumptions, and his results showed an increase in overall efficiency over a year by 4%. Also, a 2D simulation in this regard showed the possibility of increasing the working fluid's temperature to over 400°C [22]. Other efforts of this type used a three dimension model to take into account the radiation exchange by different segmented surfaces inside the receiver along the pipe's length. This study showed the hot mirror receiver effectively reduced the IR losses at a higher temperatures, reduced the thermal stress on the glass cover and suggested use in a hybrid system [23].

In this paper, the issue of the temperature restrictions imposed by the application of selective coatings on the absorber pipe, which operate close to the thermal limit of the coating, is discussed. Higher heat transfer fluid temperatures are beneficial not only to improve thermodynamic efficiency, but further allow industries which require high temperature (>700K) input to utilize the heat directly.

A hot mirror receiver unit capable of high HTF temperatures was investigated. The theoretical description is summarized from a more detailed account [23] in the second section. In the third section, the simulation based on the model is used to compare novel experimental results with theoretical expectations. The experimental data is close to theoretical expectations, with results diverging from the simulation by at most 6% at around 700K, and Chi squared test p-values around 0.99, with the worst at 0.80. The simulations underestimate the performance, indicating that a better performance was measured experimentally.

The fourth section relies on the validations of the previous section, and discusses simulation results of some currently available coatings (ITO, Gold and Silver) to predict their performance. Their performance related to conversion and energy efficiency is displayed (Figure 107 – 12).

## 2. Mathematical Model

A model was developed that describes the different heat transfer interactions inside receiver unit, allowing for a hot mirror coating on the inside of the glass cover. In this section, this model is briefly described. More detail can be found in Reference [23]. The physical basis for our model uses energy conservation for the thermal interactions in the receiver as shown in Figure 1.

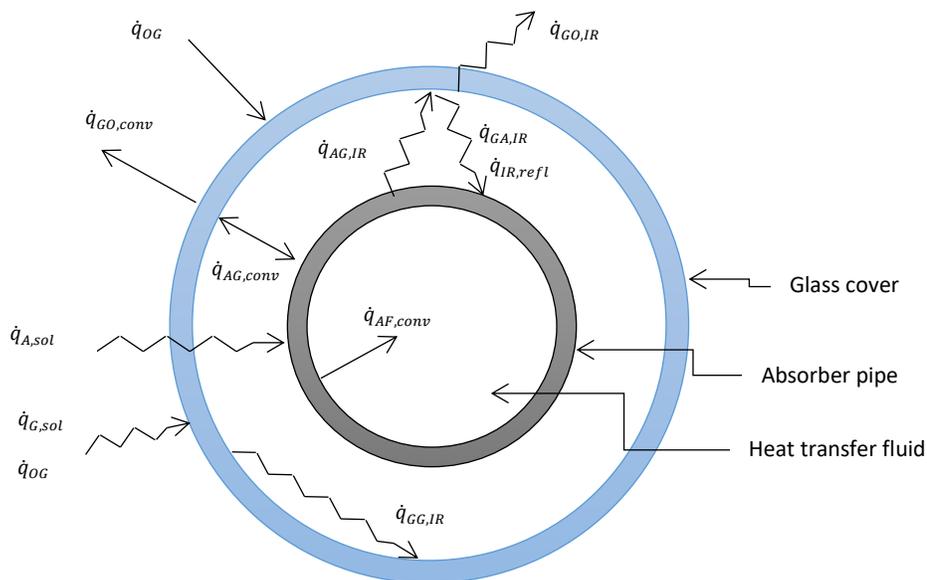

Figure 1: Heat exchanges between the glass cover, absorber pipe, heat transfer fluid and the outside environment [23].

Solar radiation is absorbed by the Glass Cover (GC), $\dot{q}_{G,sol}$ and the Absorber Pipe (AP), $\dot{q}_{A,sol}$, resulting in their temperature increase. The surfaces radiate in the IR ($\dot{q}_{AG,IR}$, $\dot{q}_{GG,IR}$, $\dot{q}_{GA,IR}$) which is either absorbed, transmitted or reflected $\dot{q}_{IR,refl}$ by the intercepting surfaces, or else lost to the outside ($\dot{q}_{GO,IR}$). Notably, the presence of a hot mirror coating on the inside glass cover results in dominant reflection of IR there. Convection heat transfer is important mode from the absorber pipe to the HTF $\dot{q}_{AF,conv}$ and from the glass cover to the outside, $\dot{q}_{GO,conv}$.

A finite volume method (FVM) is used to discretize the absorber pipe, glass cover and the Heat Transfer Fluid (HTF) into Control Volumes (CV) along the circumference and the axial direction.

An energy balance relationship of the form

$$\left(\sum \dot{q}_{COND} + \sum \dot{q}_{RAD} + \sum \dot{q}_{CONV} + \sum \dot{q}_{sol}\right)_j = 0 \qquad (1)$$

is imposed on every CV labelled on the absorber pipe and the glass cover under steady state conditions, where the net heat flux due to conduction $\dot{q}_{COND}$, radiation $\dot{q}_{RAD}$, convection $\dot{q}_{CONV}$ and solar radiation $\dot{q}_{sol}$ are defined. The heat absorbed by the HTF along a single control volume is given by the equation

$$\left(\dot{q}_{AF,conv}\right)_j = \dot{m} C_p \left(T_{j+1}^F - T_j^F\right) \qquad (2)$$

For the computation of heat conduction, the Fourier law was applied, while for convective heat transfer Newton's law of cooling was used in conjunction with the Gnielinski correlation [24]. Radiation (visible and IR) was described as emanating from every single CV onto every other visible CV resulting in surface property dependent interaction. Up to two consecutive reflections are taking into account.

Computational solutions for this model are found from our simulation code described in more detail in [23]. In section three, the experimental setup for measurement of the receiver unit performance indoors under controlled heating conditions is described. The simulation code was changed accordingly. The simulation can supply numerical values for the temperature of the HTF along the length of the pipe, the temperature profile of the absorber pipe and glass cover along the circumference and details of the heat losses resulting from the various mechanisms. Using this information, the thermal efficiency of the system can be deduced, i.e. efficiency to convert incoming solar radiation into internal energy of the HTF.

Efficiency is calculated using the ratio of heat gain in the fluid to the total incident solar energy, which takes all the possible losses into account [25]. The "local" efficiency for any CV element is

$$\eta_{CV} = \frac{\left(\dot{q}_{AF,conv}\right)_j}{\left(q_{sol}\right)_j} \qquad (3)$$

where $q_{sol}$ the incident solar flux, and $\left(\dot{q}_{AF,conv}\right)_j$ is defined by equation (2). The "integrated" receiver efficiency for some number $n$ of CV's along the system's length is

$$\eta = \frac{\sum_{j=1}^{n} \left(\dot{q}_{AF,conv}\right)_j}{\sum_{j=1}^{n} \left(q_{sol}\right)_j} \qquad (4)$$

In many applications conversion to electricity is desired. The maximum "overall" efficiency $\eta_{max}$ of the solar plant and conversion to turbine work for electricity generation can be obtained using a theoretical Carnot cycle. This is for simplicity and comparative purposes and is as below:

$$\eta_{max} = \frac{\sum_{j=1}^{n}(\dot{q}_{AF,conv})_j}{\sum_{j=1}^{n}(q_{sol})_j} \times \left(1 - \frac{T_{res}}{T_{HTF}}\right), \tag{5}$$

with $T_{res}$ as ambient used as the cold reservoir and $T_{HTF}$ as HTF temperatures respectively

## 3. Experiment and code validation

### 3.1. Experimental analysis

A receiver unit was constructed in order to validate the theory and simulation described in the previous section. The unit was constructed as it would be used in a functional PTC but was heated by heating elements situated inside the absorber tube, as opposed to from the outside by concentrated sunlight. This required us to change the simulation code slightly, in order to determine the thermal properties of the unit, and this is discussed below.

### 3.1. A. Experimental setup

The receiver unit (RU) was tested indoor. The length of the tested receiver unit was 2.7 m at 25 °C. It consisted of a mild steel absorber pipe outer/inner diameter of 3.2/2.8 cm and joined pieces of Pyrex glass cover outer/inner diameter of 5.8/5.4 cm with a length of 1.35 m each (see Figure 2 and Figure 3). Two such pieces were joined to give a total receiver unit of length 2.70 m. The Pyrex glass pieces were joined with a brass section in the centre of the Absorber Pipe (AP). The central brass piece, glass cover and the absorber pipe were vacuum insulated using flame resistant high temperature silicon. The space was evacuated using an Alcatel vacuum pump (Dual stage rotary vacuum pump input:208-230VAC, 60/50HZ). The high temperature silicon also provided some degree of thermal contact insulation.

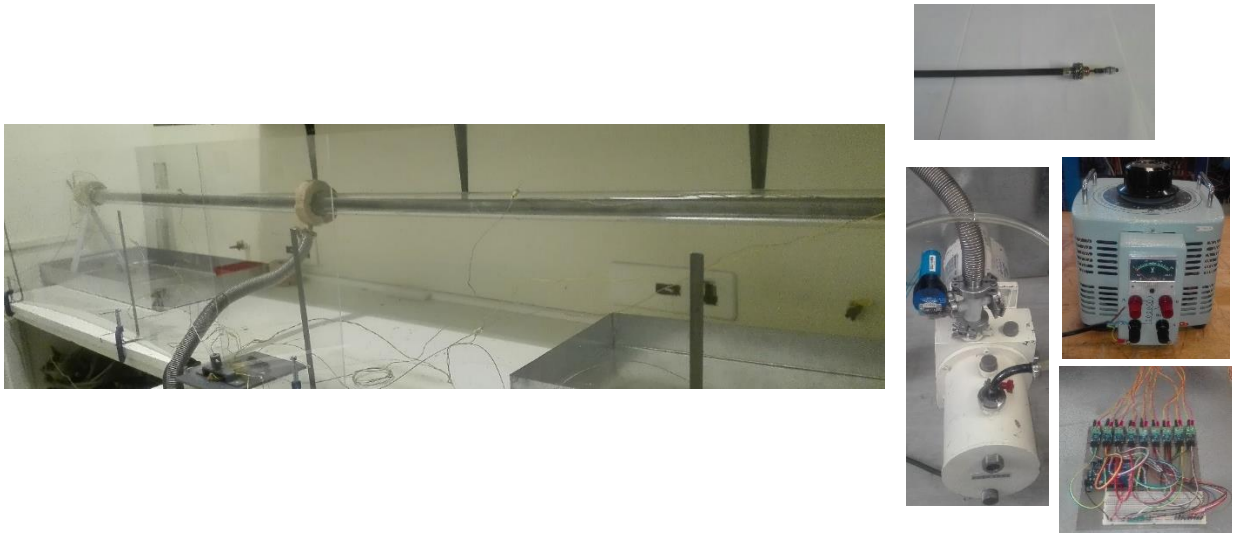

Figure 2: The receiver unit set up in the laboratory, behind a PERSPEX safety shield, on left. On right, top: the heating element. Right center: vacuum pump (left) and variac, and temperature logging unit below.

Two heating elements (1.5kW tubular element, outer diameter 8mm and length 1172mm, see Figure 2) inside the absorber pipe of the receiver unit brought the HTF temperature to the desired value. The heating element power was adjusted using a variac (Single phase variable transformer, input: 220VAC 50Hz output: 0-5kVAC and 20A, see Figure 2) and was determined by logging the current and voltage output, with an error of ±7W. The heating elements were both 1.2 m in length, with 8 mm (± 0.1 mm) outer diameter, cold resistance of 16 Ohm (±0.1 Ohm), and joined electrically inside the absorber pipe. In order to prevent the heating element from touching the absorber pipe, spacers were introduced to center the heating elements. The absorber pipe itself was filled with sand to mimic the presence of a heat transfer fluid and to distribute the heat evenly to the absorber pipe surface.

Five thermocouples (K-type thermocouples with a glass fibre twisted insulation, Nickel-Chromium alloy temperature range -200 ℃ to 1350 ℃) were mounted, two on the absorber pipe and three on the glass cover, to determine the average temperatures (±1K) and heating behavior of the receiver unit along its' length, see Figure 3. From the temperature information, the heat loss to the environment could be determined.

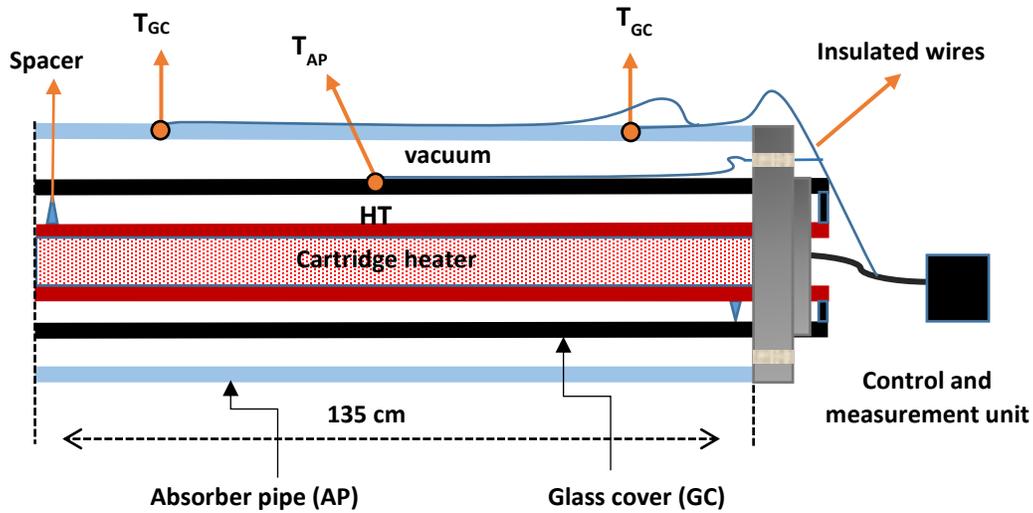

Figure 3: A section of the receiver unit. Glass cover, absorber pipe, spacers, heaters, wiring and thermocouples are shown. "T" represents the thermocouple position.

As indicated in Figure 3, the thermocouples wires were connected to the logging unit (see Figure 2), which allowed regulation and adjustment of the heater power in accordance to temperature requirements.

The experimental procedure was initiated by evacuating the space between the absorber pipe and the glass cover, down to a pressure of < 0.1 mbar (measured with a KJL 275i series vacuum gauge), which is sufficient to greatly reduce convective heat transfer in that region [26]. The ambient temperature was noted. Next, a power setting on the variac was chosen, initially around 50 W. A more accurate estimate for the electrical power to the heating elements was determined subsequently using the voltage measurements with a Brymen TBM815 voltmeter (errors on V(AC) = 0.5% & Resistance = 0.1%) from the variac and the temperature dependent resistance of the heating elements. The temperature measured by the thermocouples was noted using a dedicated microcontroller (Arduino Mega with thermocouple shield (MAX6675)) and displayed on a logging computer. The system was allowed to reach a thermal equilibrium, indicated by a stable reading of absorber pipe temperature. The time to reach equilibrium could vary up to about three hours between measurements. Once equilibrium was reached, the temperatures were noted, and the variac setting was increased to the next higher power setting (typically 50 W higher). The experiment was terminated when the power input reached approximately 2kW, due to concerns of overheating of the vacuum system.

At equilibrium, the electrical power required to maintain the absorber pipe temperature equalled the heat loss of the receiver unit through the glass cover. The temperatures along the receiver unit elements were approximately similar to within a few degrees. Heat losses were reported as Power density (Watts per meter) of receiver unit.

### 3.1. B. Adaptation of the Algorithm

The source of thermal energy for the receiver unit in the experimental setup described above was an electrical resistance heater with a constant rate of heat generation adjustable by a Variac. The heat transfer model is essentially similar to our description above, with the following modifications: the heat originates from heating elements inside the absorber pipe, raising the absorber pipe surface temperature. The HTF term $\dot{Q}_{AF,conv}$ can therefore be removed from the calculations. Further, the heating produced by the solar radiation on the absorber pipe, $\dot{q}_{A,sol}$, is equated to the heat generated by the heating element. The heating in this setup is fairly symmetric, while in a solar heated situation, the side facing the mirrors would be hotter.

Under steady state operating conditions, the absorber pipe and the outer cover (glass, hot mirror) reach a different stagnation temperature. In addition, the heat loss and the heat gain of each element in the receiver unit must equal the total rate of heat generation of the heating elements $\dot{E}_{gen}$

$$\dot{q}_{G,amb} = \dot{q}_{G,cond} = \dot{q}_{AP,G} = \dot{E}_{gen},$$

where $\dot{q}_{G,amb}$ is the rate of the heat transfer from the glass cover to the surroundings, $\dot{q}_{G,cond}$ is the conduction through the glass cover layer, $\dot{q}_{AP,G}$ is the heat transfer from absorber pipe to glass cover.

### 3.1. C. Experimental Results

Two experiments were performed as outlined above. The first used the above system and contained no coating on either the glass cover or the absorber pipe, and is designated the name "Bare". This system was investigated because it is the simplest scenario, so any applied coating should perform better in order to be considered. It provided an additional check on our simulation, and allowed us to make fine adjustments to the experimental procedure.

The second experiment used a "hot mirror" coating on the glass cover. In the first attempt, a plastic based, transparent material coated with ITO was used, similar to the material utilised as a hot mirror for house windows for interior climate control management. It was not rated for the temperatures experienced inside the receiver unit, hence deformed and partially melted. Subsequently, a thin, aluminum based metal sheet with an IR reflectivity of approximately 0.92 was used

(obtained from the MIRO SUN Alanod Solar Company datasheet). The sheet is not transparent to solar radiation, however, this was immaterial, since in the experiment the effects of IR reflectivity were tested, not of solar transparency (which was in any case absent in our simulation). The sheet served to mimic the effects of a hot mirror system. The main point of the experiment was to validate the theory and the simulation, with particular emphasis on the IR reflection component, and this must hold for any values.

**The "bare" receiver unit**

The results of the "bare" pipe experiment are displayed in Figure 4. The power density is displayed on the horizontal axis, in units of Watts per meter, and the measured and simulated temperatures are displayed on the vertical axis, in units of degrees Celsius. Both the experimental results (EXP) and the results from the simulation (SIM) are displayed.

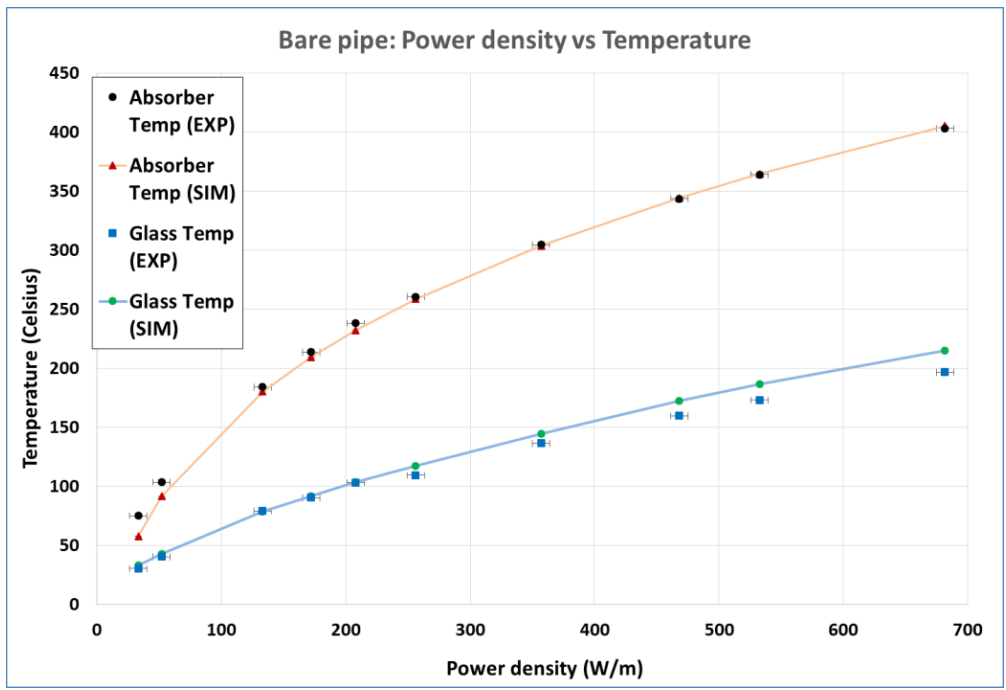

Figure 4: Experimental and simulated results for the temperature profile at different heating powers for a receiver unit without any coating, designated "bare".

A Chi squared goodness of fit gives p-values of >0.99 for the glass cover and >0.995 for the absorber pipe.

**The receiver unit with "hot mirror" coating**

The results of the second experiment and the simulation are displayed in Figure 5. The axes display the same units as for the "bare" case. It is seen that the experimental (EXP) and simulated (SIM) results for the glass cover diverge at most by 3%, while those of the absorber pipe by at most 6% at around 500 C°, where the simulation underestimates the temperature.

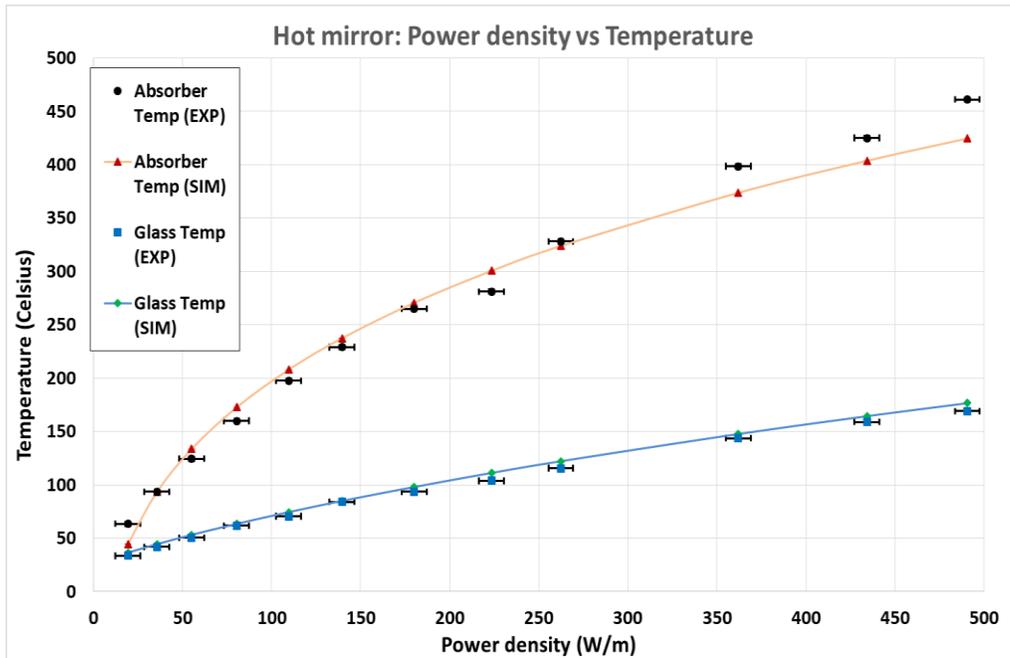

Figure 5: Simulated and experimental results of receiver unit coated with a hot mirror at various temperatures.

A Chi squared gives p-values of >0.995 for the glass cover. For the absorber pipe, a p-value of >0.80 was obtained, and the point at 44°C was considered an outlier so only 10 degrees of freedom were included. The main contribution for the divergence comes from high temperature points, but the simulation underestimates experimental performance. This divergence is likely due to temperature dependent simulation parameters, and can be addressed in a more accurate model. The effect of the coating can be seen in Figure 6, where the experimental data from the bare pipe and the hot mirror coated pipe are displayed.

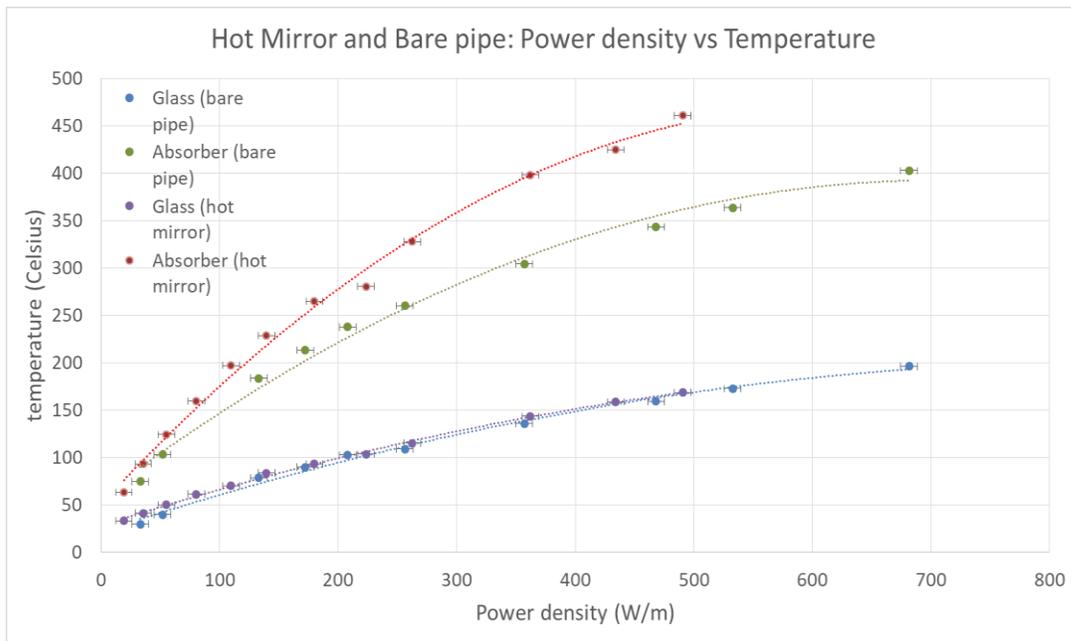

Figure 6: Comparison between experimental results from bare pipe and hot mirror receiver. The hot mirror absorber pipe is capable of reaching much higher temperatures at the same power input.

The absorber pipe in the hot mirror case is significantly hotter, indicating better thermal retention, and would therefore have a better capability of heating the heat transfer fluid inside to high temperatures. The glass temperature in both cases is similar, as expected.

### 3.1. D. Simulation of realistic scenarios

The validation results of our simulation shown in the previous section as well as in [23] [27] allows comparison of the performance of more realistic coatings and associated efficiencies, as well as obtain thermal profiles and heat losses of the receiver unit. Our original simulation from [23] was used in conjunction with a non-uniform solar heat flux distribution as suggested by Jeter [28].

The simulation made use of the same operating conditions and design parameters that were used to simulate the SEGS (Solar Electric Generating System) LS2, which is one of three generations of parabolic troughs installed in the nine SEGS power plants in California [29]. The performance of various possible hot mirror coatings on the glass cover (ITO, gold, silver) were compared. The bare case was also simulated as a reference. The simulation parameters are shown in Table 1.

Table 1: Parabolic trough solar collector parameters used in the simulation.

| Parameter | Value |
|---|---|
| Collector aperture (W) | 5 m |
| Focal distance (f) | 1.84 m |
| Absorber internal diameter ($D_{a,inner}$) | 0.066 m |
| Absorber external diameter ($D_{a,ext.}$) | 0.070 m |
| Absorber emissivity (IR) | 0.15 |
| Glass internal diameter ($D_{g,inner}$) | 0.109 m |
| Glass external diameter ($D_{g,ext.}$) | 0.115 m |
| Parabola specular reflectance ($\rho_p$) | 0.93 |
| Incident angle | 0.0 |
| Solar irradiance | 933.7 W/m$^2$ |
| HTF | Natural Salts |
| Mass flow rate (kg/sec) | 5 kg/s |
| Heat capacity of HTF | 1530J/(kg K) |
| Temperature HTF (inlet) | 375.35 K |
| Temperature Ambient | 294.35 K |
| Wind speed | 1.5m/s to 2.6 m/s |

Table 2 shows the glass cover and absorber pipe optical properties for all the scenarios considered in the simulation. The absorber pipe surface in the bare case has no selective optical properties. It is assumed gray in its spectral characteristics.

Table 2: Simulated Scenarios, optical characteristics of the GC and AP for different hot mirror materials.

| | | Bare | | ITO | | Gold | | Silver | |
|---|---|---|---|---|---|---|---|---|---|
| | | GC | AP | GC | AP | GC | AP | GC | AP |
| Visible | Transmissivity | 0.935 | 0 | 0.875 | 0 | 0.44 | 0 | 0.4 | 0 |
| | Reflectivity | 0.04 | 0.14 | 0.1 | 0.14 | 0.529 | 0.14 | 0.568 | 0.14 |
| IR | Transmissivity | 0 | 0 | 0 | 0 | 0 | 0 | 0 | 0 |
| | Reflectivity | 0.14 | 0.14 | 0.85 | 0.14 | 0.78 | 0.14 | 0.95 | 0.14 |

Infrared (IR) reflecting materials play an important role in solar efficiency. They generally fall in three categories [21] [16].

*1. Thin metal layers:*

A good IR reflector can be made by depositing on the glass substrate a thin metal film, made of a metal with a low IR emissivity, such as silver, gold, copper, or aluminum. However, to obtain a good IR reflectivity (>80%), the solar transmissivity of a simple metal layer becomes quite limited (less than 40%). Silver has a higher solar transmissivity but not as stable to the environment as Gold. In our simulation, optical properties for a metal layer of thickness 200 nm was chosen [30].

*2. Doped semiconductors with an appropriate band gap:*

Compared to metal films, doped semiconductor thin layers have usually a lower IR-reflective coefficient (usually less than 85%) but they have a better transparency for the visible light (up to 80%). A 200nm film of Indium Tin Oxide ($In_2O_3$:Sn) has a solar transmissivity of about 80%, with an IR-reflective coefficient of 75%. ITO is frequently most applied because of its lower deposition temperatures and high solar transmissivity.

*3. Composite layers:*

The solar transmissivity of a metal thin film can be increased using a material with high refractive index as an antireflective layer. Composite layers are usually made by a thin noble metal layer between transparent dielectric layers. The solar transmissivity can be increased above 70%, but the realization is more difficult, since different layers of accurately controlled thickness must be deposited. For this reason, this solution is probably more expensive and technically difficult. Three materials, ITO, Gold and Silver as hot mirror materials are chosen in the simulations below

## 4. Simulation Results

### A. Comparative performance of hot mirror coatings

Simulations were constructed for the different coating materials: bare, ITO, Gold and Silver subject to the parameters of Table 1 and Table 2. The question of how the different coatings affect a variety of parameters such as the thermal distribution of the HTF along the length, the thermal distribution of both the absorber pipe and the glass cover along the circumference and related efficiencies, is addressed.

The results for the absorber pipe temperature distribution along the length and circumference are shown in Figure 7. The absorber pipe temperature for an angle of $180°$ is shown in the top graph. The angles appearing in the figures along the circumference below are defined for the cross section of the glass cover and absorber pipe, with an angle of 180º facing the center of the mirror, and an angle of 0º or 360º facing the sun. The ITO coating performs best in terms of temperature increase, indicating best heat retention and visible transparency. The bare pipe scenario does surprisingly well, and the gold and silver coatings only manage to surpass it at a later stage, which will be explained later.

Along the circumference, the temperature gradient at a randomly chosen length of 100 m is highest for bare and ITO. In the bare case, it is due to lack of heat being spread by any hot mirror coating, and the distribution can only occur via conduction along the pipe material. For ITO, the higher temperature gradient it is due to the more elevated temperature of the absorber pipe at that length compared to the other materials.

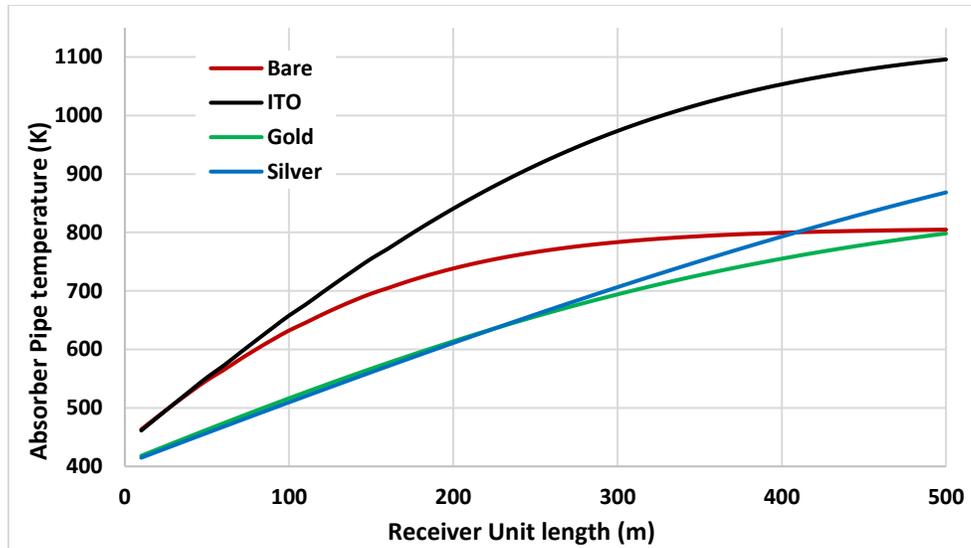

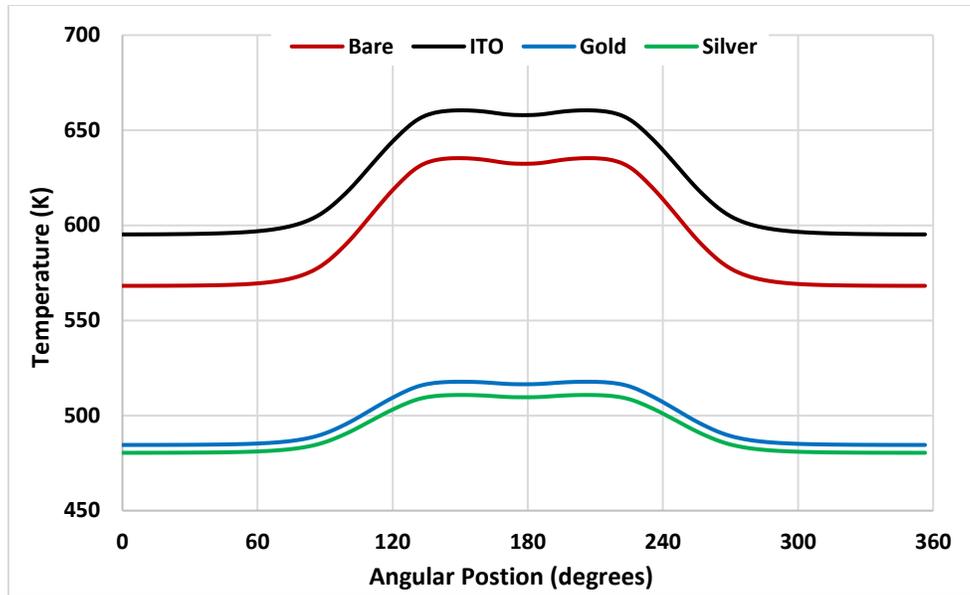

Figure 7: (Top) The absorber pipe temperature at an angle of 180° (facing the center of the parabolic mirror) along the length of the receiver unit. (Bottom) Temperature profile around the circumference of the receiver unit absorber pipe at a length of 100 m.

There is a temperature gradient of around 50 K or lower present in all cases, which is close to the suggested maximum. The temperature distribution around the circumference is similar in shape to the suggested incoming radiation profile that enters the receiver unit via parabolic trough collector mirror [28].

The results for the thermal profiles of the glass cover for the four different materials along the length and the circumference of the receiver unit are shown in Figure 8. The glass cover temperature is highest in the bare case below 300 m due to the glass cover being completely unprotected from IR absorption. The ITO case glass cover temperature dominates after 350 m due to the higher temperature (~300 K higher) of its absorber pipe compared to all other cases.

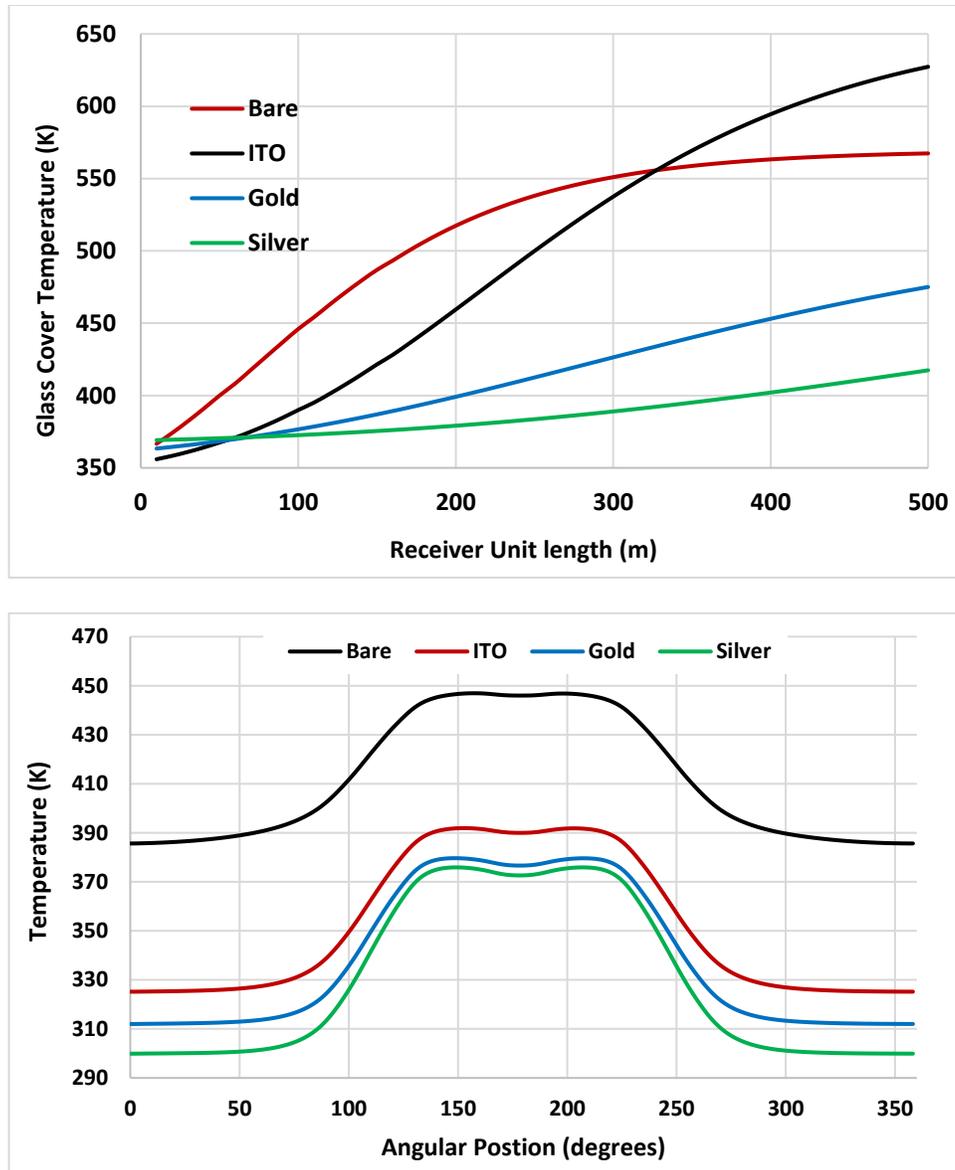

Figure 8: (Top) The glass cover temperature at an angle of 180° (facing the center of the parabolic mirror) along the length of the receiver unit. (Bottom)Temperature profile around the circumference of the receiver unit glass cover at a length of 100 m.

The glass cover temperature experiences large thermal gradients of approx.50 K around their circumference, but their absolute temperature is not excessive. A smaller thermal gradient is desirable since it induces fewer thermal stresses in the glass, allowing a longer lifetime. The low glass cover temperature will allow coatings without danger of thermal decomposition even at high absorber pipe temperatures. Even for the highest absorber pipe temperature reached in the ITO case in excess of 1000 K, the glass cover temperature does not exceed 570 K. The ITO coating can withstand temperatures over 700 K [31] highlighting a major advantage of hot mirror coating relative to the selective absorber coating placed on the absorber pipe.

Figure 9 compares the results for the HTF temperature variations along length of the receiver for the different coatings. The temperature profile behavior is by necessity very close to the top graph of Figure 7, since the absorber pipe heats the HTF via conduction. ITO outperforms all other considered coatings.

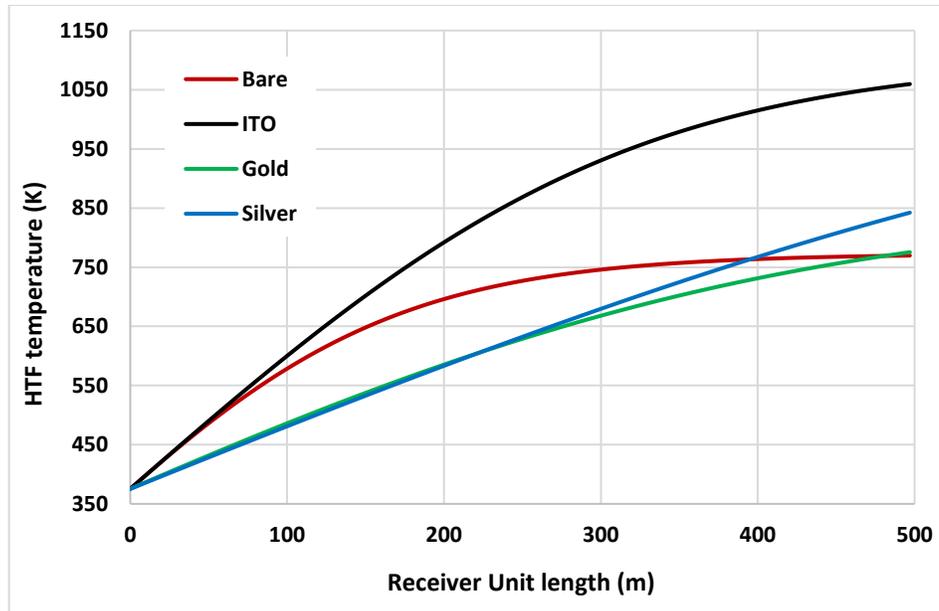

Figure 9: HTF temperature distribution along the length of the receiver unit for the four materials.

Figure 10 displays the efficiency of the materials to retain heat (defined by Equation 4), which is displayed in the top graph, and, by inclusion of Carnot efficiency, the plants' maximum ability to convert heat into electricty (defiend by Equation 5). These graphs are main results. The integrated efficieny is a measure of ability of heat conversion. The bare case starts out incrementally more efficient than all others due to its unimpeded visible transmissivity, but is rapidly overtaken by ITO, which dominates the remainder of the graph. Gold and silver coatings only start to dominate the bare case scenario after 400 meters.

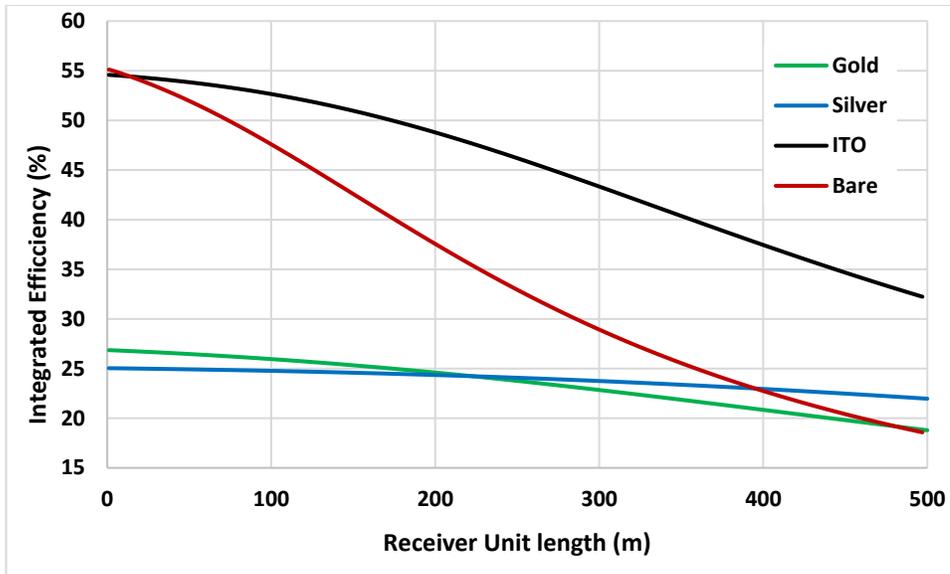

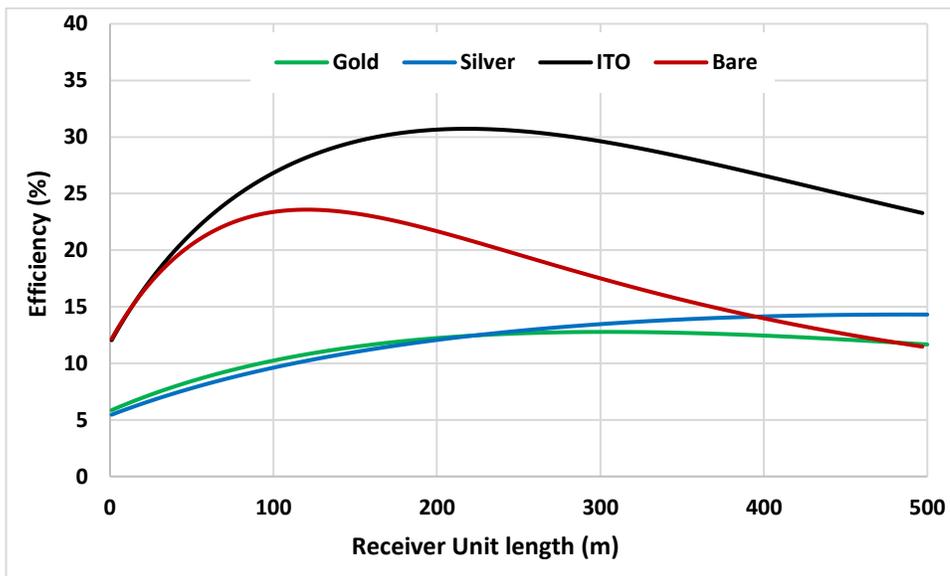

Figure 10: (Top) integrated efficiency, essentially displaying the thermal retention efficiency of the receiver unit. (Bottom) Total efficiency. The maxima indicate the maximum attainable efficiency of hot to electricity conversion.

The bottom graph of Figure 10 is a measure of the overall performance of the solar plant in converting sunlight into electricity. The graphs were obtained using Equation 4 and Equation 5. The maxima of the curves indicate the highest attainable efficiency of the solar plants to convert solar energy into electricity, as well as the optimum length at which this can be achieved. ITO reaches the highest plant efficiency of ~31%, while the bare pipe reaches 26%. The Gold and Silver coating peak around 12% efficiency, signaling weak performance. The likely causes for this ranking will be discussed in the next section.

**B. Effects of transparency and reflectivity**

Transparency to visible radiation allows more energy to fall onto the pipe, while reflectivity of the glass cover in the IR prevents more thermal radiation losses. In the case of ITO, Gold and Silver, materials with high reflectivity in the IR often have reduced transparency in the visible (see Table 2). Accordingly, the performance of two hypothetical materials, H1 and H2 were selected for a comparative study. The material H1 has high IR reflectivity but relatively poorer visible light transparency, while these properties are reversed for H2. The main question to be answered is whether high visible transmissivity is a preferred characteristic to high IR reflectivity as a general guideline for material design. The glass cover spectral characteristics are shown in Table 3 and for the absorber pipe similar to Table 2.

Table 3: Optical properties on the GC for H1 and H2.

|         |                | H1    | H2    |
|---------|----------------|-------|-------|
| Visible | Transmissivity | 0.875 | 0.95  |
|         | Reflectivity   | 0.1   | 0.027 |
| IR      | Transmissivity | 0     | 0     |
|         | Reflectivity   | 0.95  | 0.85  |

The broad patterns of the behavior may be deduced from the simulations. The glass cover temperature for H1 (strong IR mirror) has a smaller thermal gradient and absolute temperature than H2. The absorber pipe temperature along the length is very similar and only diverges around 950 K, with H1 becoming hotter. There is a similar temperature gradient present in both cases. Figure 11 shows the integrated efficiency of the materials H1 and H2. H2 dominates initially, due to its visible transparency, but H1 overtakes at higher temperatures. This seems to be a pattern observable also for the realistic materials studied above. In such cases, where different materials dominate in different temperature ranges, the designer could consider a hybrid plant, using the optimal material for each length.

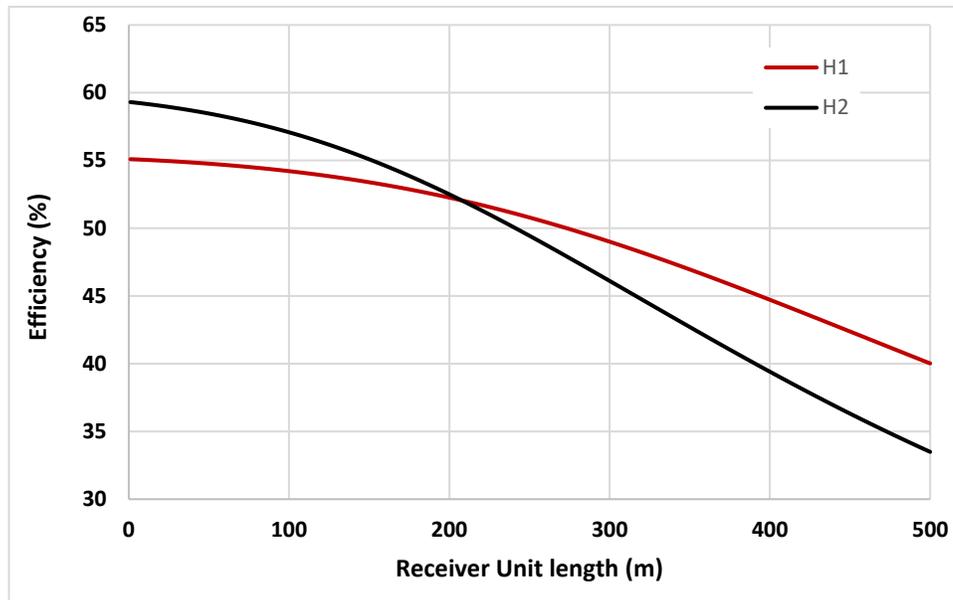

Figure 11: The integrated efficiency of the two hypothetical materials, H1 and H2, is shown as a function of receiver unit length.

**C. Selective coating and hot mirrors**

In this section, the behavior of selective absorber coating (herein designated SC) on the absorber pipe and hot mirror coating on the glass cover perform comparatively were determined. It is natural to ask if a combination of selective and hot mirror coating, used simultaneously, can aid performance. Three scenarios are considered, one of selective coating only (SC), one of a selective coating on the absorber pipe and hot mirror coating on the glass cover (SC+HM) and one of the previously discussed hot mirror coating scenarios. The simulation values are given in the Table 4.

Table 4: Spectral Specifications for the three scenarios, SC, SC+HM and HM, for the absorber pipe and glass cover.

|    |         |                | SC    | SC+HM | HM    |
|----|---------|----------------|-------|-------|-------|
| GC | visible | Transmissivity | 0.935 | 0.875 | 0.875 |
|    |         | Reflectivity   | 0.04  | 0.1   | 0.1   |
|    | IR      | Transmissivity | 0     | 0     | 0     |
|    |         | Reflectivity   | 0.14  | 0.85  | 0.85  |
| AP | visible | Transmissivity | 0     | 0     | 0     |
|    |         | Reflectivity   | 0.04  | 0.04  | 0.04  |
|    | IR      | Transmissivity | 0     | 0     | 0     |
|    |         | Reflectivity   | 0.85  | 0.85  | 0.14  |

The HM coating reaches the lowest stagnation temperature, while there is very little difference in the temperature distributions of the SC and the SC+HM. The SC+HM scenario has the disadvantage of the relative low temperature ceiling on the absorber pipe imposed by SC coating thermal breakdown.

For the integrated efficiency, Figure 12, a modern SC will outperform an ITO coating as well as a combination (SC+HM). At elevated temperatures, the combination may perform slightly better, but then the SC temperature ceiling would be exceeded. As before, the HM advantage still lies with its much higher temperature ceiling, making it a choice for application where high temperature of the HTF is required without conversion to electricity.

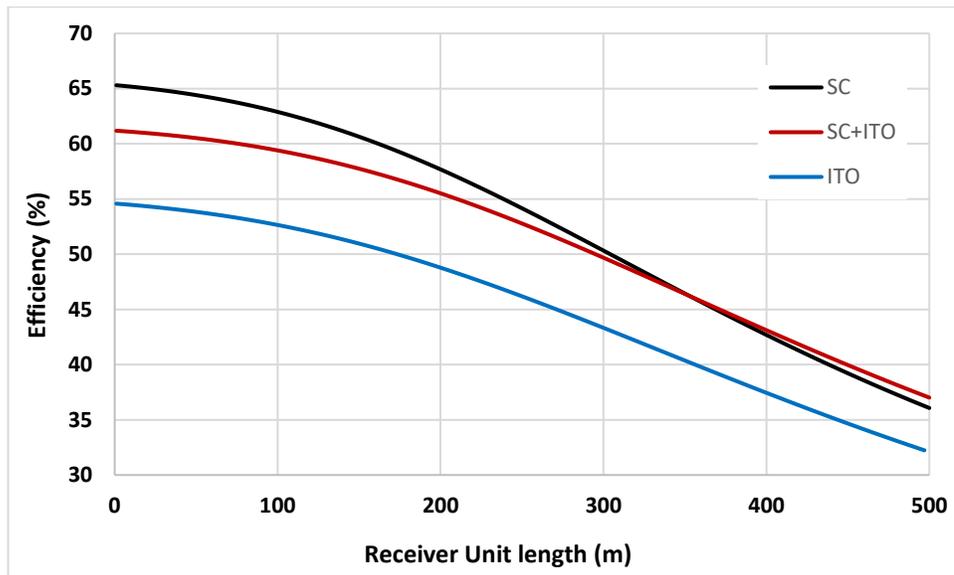

Figure 12: Graph showing integrated efficiency of SC, SC+HM and HM as a function of receiver unit length.

For maximum total plant efficiency, the SC peaks at 37% but does not reach it due to temperature limitations. The HM coating reaches its peak of 31%. Nevertheless, in this case, even the terminated SC efficiency peak (at ~34%) surpasses the HM maximum, indicating that it is still a better choice in terms of efficiency.

## 4. Conclusion

The experimental results from the bare and hot mirror coated receiver unit indicated that the hot mirror system was capable of greater heat retention, which was evident by the higher temperature that was reached, around 100ºC at 500 W/m power density. Further, the Chi squared comparison between the experimental and simulated values, which yielded p-values of 0.99 for glass cover and 0.995 for absorber pipe on the bare pipe, and 0.995 for glass cover and 0.80 for absorber pipe on the hot mirror coating, indicated that the model is reasonable accurate to describe a hot mirror coated receiver unit. The simulation underestimated the performance of the hot mirror at higher temperatures.

Comparing simulations for other hot mirror coating candidates (ITO, silver and gold) suggested that ITO can reach very high stagnation temperatures (>1200K). Silver and Gold coatings performed poorer even than a bare pipe initially, due to their reduced visible light transmissivity.

The effects of IR reflectivity and visible transmissivity were isolated and studied in a hypothetical case. While the coatings showed similar behavior to real materials in terms of temperature distribution, the visible transmissivity was the more important characteristic in terms of efficiency, while IR reflectivity was more important in terms of a higher temperature ceiling of the HTF.

Lastly, the performance of the ITO coating against a conventional selective coating (SC), and a selective coating + hot mirror combination was compared. The optical characteristics of the SC were still superior, and the SC performed better in terms of efficiency. A combination of hot mirror and SC in the same receiver did not yield any obvious advantages. The hot mirror by itself did not outperform the SC efficiency-wise, but still was shown to have the advantage of being able to achieve much higher HTF temperatures. Even including the cut-off temperature limitations, the SC managed to reach a higher efficiency than the peak efficiency of the hot mirror considered.

The comparisons of the materials used in this paper suggest that the SC is a better candidate for high efficiency solar plants than the hot mirror coatings which were considered, while hot mirror coatings can be used for high HTF temperature applications.

In future work, the possibility of a hybrid plant could be considered, where sections are composed of different types of coatings suitable for their temperature range. In addition, hybrid hot mirror materials will be studied, whose properties are equal or superior to existing SC's, and which good candidates for high temperature and efficiency applications are. Finally, our most focused efforts are directed towards a cavity type system, which shows great promise.

**Acknowledgements**

The authors would like to thank the following entities for their kind support: MERG (Materials for Energy Research Group) and MITP (Mandelstam Institute for Theoretical Research), at the University of the Witwatersrand.